# Earth Radioactivity Measurements with a Deep Ocean Anti-neutrino Observatory

S.T. Dye[1,2], E. Guillian[1], J.G. Learned[1], J. Maricic[1], S. Matsuno[1], S. Pakvasa[1], G. Varner[1], M. Wilcox[1]

[1]*Department of Physics and Astronomy, University of Hawaii, 2505Correa Road, Honolulu, Hawaii 96822 U.S.A.*

[2]*College of Natural Sciences, Hawaii Pacific University, 45-045 Kamehameha Highway, Kaneohe, Hawaii 96744 U.S.A.*

**Abstract.** We consider the detector size, location, depth, background, and radio-purity required of a mid-Pacific deep-ocean instrument to accomplish the twin goals of making a definitive measurement of the electron anti-neutrino flux due to uranium and thorium decays from Earth's mantle and core, and of testing the hypothesis for a natural nuclear reactor at the core of Earth. We take the experience with the KamLAND detector in Japan as our baseline for sensitivity and background estimates. We conclude that an instrument adequate to accomplish these tasks should have an exposure of at least 10 kilotonne-years (kT-y), should be placed at least at 4 km depth, may be located close to the Hawaiian Islands (no significant background from them), and should aim for KamLAND radio-purity levels, except for radon where it should be improved by a factor of at least 100. With an exposure of 10 kT-y we should achieve a 25% measurement of the flux of U/Th neutrinos from the mantle plus core. Exposure at multiple ocean locations for testing lateral heterogeneity is possible.

# 1. Introduction

This report furnishes recommendations for the size and sensitivity needed by a deep ocean anti-neutrino detector near Hawaii (Hawaii Anti-Neutrino Observatory- Hanohano) to perform geophysics measurements of Earth radioactivity. The design and experience of the KamLAND project in Japan provides an excellent guide to detector size and location needed to approach two geophysics goals. This detector would be more than an order of magnitude larger than KamLAND with appropriate modifications and adaptations for operation in the deep ocean. Event rates quoted for this detector are based on an exposure of 10 kilotonne-years (kT-y) of KamLAND scintillating oil, which contains some 8.5 x $10^{31}$ free protons per kilotonne.

## 1.1 Measurement of Geo-neutrinos

The first geophysics goal is measurement, not merely detection, of the electron anti-neutrino flux from the mantle and core of Earth due to uranium and thorium (U/Th) decays. Because the concentrations of U/Th are much higher in the continental crust than in the oceanic crust and mantle, locations far from the continental crust, like Hawaii, are well suited to this measurement. Although geologists usually predict the core to be free of U/Th, the flux measurement described herein is not sensitive to electron anti-neutrino direction and therefore does not differentiate between the mantle and core. In subsequent descriptions of measurements of electron anti-neutrino flux from U/Th decays mantle refers to mantle plus core.

Whereas the concentrations of U/Th in the outermost Earth's crust can be sampled directly, measurement of the "geo-neutrino" flux provides the only viable method for determining these concentrations in the mantle. These values are known poorly at present and speculated upon by geologists. Concentrations are typically inferred from the U/Th concentrations in meteorites plus assumptions about Earth accretion and differentiation. Note that geologists generally quote such numbers without error ranges, simply because there are too many unknowns: U/Th concentrations and distributions are generally acknowledged to be informed guesses. Hence the geological community welcomes information carried by geo-neutrinos from the otherwise inaccessible inner Earth. This was made evident by the reception of the first KamLAND results on the measurement of geo-neutrinos (mostly from the local crust in Japan), published as a cover article in the 28 July 2005 issue of Nature (Araki et al., 2005a). This paper heralding the first positive detection of Earth's total radioactivity marks a beginning for neutrino geophysics, long a tantalizing goal (Eder 1966; Avilez et al., 1981; Krauss et al., 1984). The article reported 28 geo-neutrino events above background from an exposure of about 1 kT-y. Seven of these events can be attributed to the mantle (Enomoto 2005). While a start, the report does not add much new information about Earth's composition or radiogenic heat.

Several groups have made calculations of geo-neutrino fluxes (Raghavan et al., 1998; Rothschild et al., 1998; Fiorentini et al., 2004; Enomoto et al., 2005), and there are two PhD dissertations from KamLAND which contain significant modeling (Tolich 2005; Enomoto 2005).

**1.2 Search for Hypothetical Geo-reactor**

The second goal is a definitive search for a hypothetical nuclear reactor at Earth's core. This theory (Herndon 1996; Hollenbach and Herndon 2001) has not met wide acceptance by the geological community, who have generally preferred the idea that much of the U/Th rose from the molten, early inner Earth as slag, rather than sank to the core as elemental metal. Yet, many geologists say that there really is no evidence against the hypothesis since the conditions at Earth's formation are little known. Moreover, there are peculiarities in the isotopic content of Earth, and most particularly the observed high ratio of $^3$He/$^4$He coming out of oceanic volcanic hot spots (such as Hawaii and Iceland), which a natural reactor could explain ($^3$He would come from tritium decay, made abundantly in reactors).

As discussed elsewhere (Raghavan 2002), this hypothetical energy source in the range of 1 to 10 terawatts of thermal (TW$_t$) power could be the enigmatic power source driving the deep Earth plumes, and hence ultimately responsible for the motion of landmasses (plate tectonics) as well as Earth's magnetic field (geo-dynamo). The neutrino flux from this putative geo-reactor is very hard to measure at locations anywhere near electrical power reactors, especially in places such as Japan, Europe and North America (Raghavan 2002; Domogatski et al., 2004).

At KamLAND the geo-reactor would present a flux of only a few percent of that due to power reactors around Japan. This is very hard to distinguish from the power reactor flux because the energy spectrum at the source of natural or man-made reactors is essentially indistinguishable. However, since reactors at distances of a few hundred kilometers are not so far as to have all neutrino oscillations effects (see next section) washed out, a study has been made seeking an unchanging geo-reactor signal added to the power reactor flux with a time varying spectrum (Maricic 2005). This is sufficient for claiming an upper limit on such a power source (<20 $TW_t$). Although tantalizing, the study lacks the sensitivity required to detect and measure the power of the geo-reactor if it exists.

## 1.3 Neutrino Oscillations

In the following discussion, all processes that involve neutrino production and subsequent detection assume that neutrino oscillations occur, as is now established. The oscillation parameters employed are the best-fit values from global fits to all solar and reactor neutrino experiments as of this time (Araki et al., 2005b). Since the baseline of neutrino propagation considered in this context (thousands of kilometers) is much larger than the oscillation lengths (<100 km) for the energy scale under consideration, the neutrinos can be considered to a good approximation to be fully mixed. The effect of oscillations can be accounted for by reducing the event rate by a factor of

$$P(\nu_e \to \nu_e) = 1 - \tfrac{1}{2}\{\cos^4(\theta_{13})\sin^2(2\theta_{12}) + \sin^2(2\theta_{13})\} \approx 0.6 \qquad (1)$$

compared to the rate without oscillations, using current experimental values for the mixing angles.

## 1.4 Electron Anti-neutrino Detection and Analysis Windows

Electron anti-neutrinos are observed by the detection of positrons and neutrons produced by inverse neutron decays in scintillating liquid by the standard technique. The positron produces a prompt signal boosted by positron-electron annihilation with a visible energy in the detector ~0.8 MeV less than the electron anti-neutrino energy. A delayed signal at 2.2 MeV of visible energy from the formation of deuterium tags the neutron capture.

The threshold energy for inverse neutron decay is 1.8 MeV. This sets the lower bound of the geo-neutrino analysis at 1.7 MeV, adjusted for detector energy resolution. Geo-neutrino energies extend up to 3.4 MeV, which sets the upper bound for the analysis. The window for the geo-reactor analysis is 3.4 MeV to 9.3 MeV.

## 2. Geo-neutrino Detection Sensitivity

Hanohano's location in the middle of the Pacific Ocean makes it sensitive primarily to geo-neutrinos originating from Earth's mantle and core. The nominally expected event rate of geo-neutrinos of mantle origin based upon the Bulk Silicate Earth (BSE) model (Fiorentini et al., 2004) is 79 events per 10 kT-y. This is more than 2.5 times larger than the 31 events per 10 kT-y for geo-neutrinos from the oceanic and continental crusts. Because of the uncertainty of the modeling we do not assign an error estimate to the event rate from the mantle.

A geo-neutrino flux dominated by the mantle at sites near Hawaii is noted by several authors (Rothschild et al., 1998; Pakvasa 2005; Mantovani et al., 2004; Enomoto et al., 2005). The situation is reversed at a continental location with the same flux from the mantle but about eight times the flux from the crust. In this analysis the signal is the geo-neutrino event rate from the mantle. Geo-neutrinos from the crust are considered part of the background. The conclusion is that the flux of geo-neutrinos from the mantle is extremely difficult to measure at continental locations, which yield mainly a flux from the crust.

### 2.1 Geo-neutrino Background

Expected background for geo-neutrinos (based upon KamLAND experience) includes:
- $^9$Li produced by cosmic rays traversing the detector,
- Fast neutrons from cosmic rays passing near the detector,
- $\alpha$ decay of $^{210}$Po followed by $^{13}$C($\alpha$,n)$^{16}$O in the scintillating oil,
- Accidental or random coincidences,
- Anti-neutrinos from commercial nuclear reactors, and

- Anti-neutrinos from a geo-reactor if it exists.

The lithium background is due to cosmic ray muons traversing the detector, decreasing with increasing depth. At the depth of KamLAND, equivalent to ~2.1 km of water (Mei 2006), it is a major nuisance. By 4 km depth the Li background is almost negligible. The low lithium background level in KamLAND is achieved at the cost of applying tight cuts around reconstructed muon tracks, which results in the removal of a significant amount of good data. These cuts, moreover, introduce systematic errors that obscure the signal. For these reasons, the most favorable strategy is to go as deep as possible so that the cosmic ray background rate is so low that the application of cuts to remove $^9$Li events becomes unnecessary. Greater depth alleviates a multiplicity of background problems, including entering fast neutrons and significant dead time around muon transits. Background contribution due to fast neutrons is negligible in KamLAND and is ignored in this analysis. Since fast neutrons occur at the edge of the fiducial (inner, software-defined) volume they are efficiently removed in the data analysis.

We determine that overburden-dependent background is reduced to a comfortable level by a depth of 4 km. Fortunately the abyssal plane of the ocean is 4-5 km in depth. A potential site 34 km west of the Big Island of Hawaii (19.72N, 156.32W) at about 4.5 km depth meets our requirements.

The polonium background, due to alphas which interact with $^{13}$C, stems mostly from radon contamination at KamLAND. The mine levels of radon are 40 times those in the free air outside. In the course of experimental preparation the KamLAND scintillating oil was possibly exposed to mine air. While the radon itself decays in a matter of months, further decay products lead to the initially unrecognized polonium background, which can simulate inverse beta decay. The background level used for this analysis assumes a concentration of $^{210}$Po 1/100 of that at KamLAND, which is conservatively high (Suzuki).

We define "accidental" background to be due to random coincidences. Radioactivity of the detector itself contributes. Some of this radioactivity comes from the periphery of the detector, the balloon and supporting ropes in KamLAND. Thus this scales with detector outer surface area not volume. For present purposes we take the conservative assumption that the rate per unit volume will be the same as in KamLAND, but we can doubtless do better than indicated.

The calculation of the contribution from distant commercial power reactors can be carried out to about 2% precision, and should be well known. A detector near Hawaii would record about 12 events per 10 kT-y exposure. Locations in the southern ocean and near Australia realize contributions lower by about a factor of two, whereas contributions at continental locations in the northern hemisphere are typically higher by at least an order of magnitude (Rothschild et al., 1998). Clearly, Hawaii has a very low rate compared to other possible locations making it extremely suitable for measuring geo-neutrinos from the mantle. Our estimates include the small contribution due to long-lived reactor products. The contribution from nuclear powered ships and submarines warrants consideration (Detwiler et al., 2002). Submarine power plants are in the range of 100 $MW_t$, as compared to 2 $GW_t$ for power reactors. Ships cruise at a small fraction of maximal power generally, and are shut down in port at Pearl Harbor. Although not included in our background estimates, these could be accounted for with cooperation of the military (since we only need to know flux at the detector and not power or range of the ships).

In arriving at detector parameters for geo-neutrino measurement, we assume that the geo-reactor power is zero. Were it to exist, then we would know the power quite well from the measurements above 3.4 MeV of neutrino energy. An Earth-centered geo-reactor would contribute 19 events per $TW_t$ per 10 kT-y to the geo-neutrino measurement.

The final "background" is the contribution to the U/Th neutrino flux from the oceanic and distant continental crusts. An uncertainty of about 20% is assigned. This is consistent with geological models (McDonough and Sun, 1995) and the detailed studies for the KamLAND site (Enomoto 2005).

We have made a preliminary estimate of the additional flux of neutrinos due to the proximity of the Hawaiian Islands. When we assume that the Big Island can be modeled as a cone of mid-ocean ridge basalt of height 10 km and radius 100 km, and that our detector is located a mere 10 km from the effective source center, the contribution would amount to only 5% of the mantle flux. When a site is finally chosen we will have to do a more careful calculation, but for now we can safely conclude that proximity to the Hawaiian Islands will not affect the experimental goal of measuring the mantle flux.

**2.2 Geo-neutrino Signal Analysis**

Table 1 presents the numbers of events expected for the geo-neutrino signal analysis. The total geo-neutrino "background" rate at 4 km depth (without geo-reactor but including neutrinos from commercial reactors and the crusts) is 96 ± 7 per 10 kT-y, compared to a mantle signal rate of 79 per 10 kT-y. The background subtracted mantle geo-neutrino signal would be 79 ± 20, a 25% measurement on the total rate alone (not using spectrum).

In order to confirm the above conclusion, we performed simulations where the combined energy spectrum of the signal and background are varied randomly and a multi-component fit is done for the number of signal events. The results improve, particularly depending upon how well we are able to constrain the background components.

**2.3 Advantage of Oceanic Site**

The advantage of an oceanic site for measuring the geo-neutrino signal from the mantle and thereby Earth radioactivity is demonstrated by the following example. We consider equal 10 kT-y exposures for several potential geo-neutrino detectors including Borexino (Giammarchi and Miramonti), SNO+ (Chen), and Hanohano. Assuming lateral homogeneity in the mantle each detector would record a signal of 79 events. Background event numbers are estimated from the depth and geographic location of each detector and an assumed level of $^{210}$Po radio-purity 100 times better than reported by KamLAND (Araki et al., 2005). Table 1 presents the numbers of events from each source along with their uncertainties expected in the detectors considered.

Table 1: The numbers of events expected for the mantle geo-neutrino analysis for energies between 1.7 and 3.4 MeV. We assume the $^{210}$Po background to be 100 times purer than the level for the scintillating oil at KamLAND and the reactor background to be known to 4%.

|  | Events (10 kT-y)$^{-1}$ | | |
| --- | --- | --- | --- |
|  | **SNO+** | **Borexino** | **Hanohano** |
| $^9$Li | 0 ± 0 | 3 ± 1 | 3 ± 1 |
| $^{210}$Po | 8 ± 2 | 8 ± 2 | 8 ± 2 |
| Accidental | 42 ± 1 | 42 ± 1 | 42 ± 1 |
| Reactor | 528 ± 21 | 295 ± 12 | 12 ± 1 |
| Crust Geo-νs | 368 ± 74 | 279 ± 56 | 31 ± 6 |
| **Background** | 946 ± 77 | 627 ± 57 | 96 ± 7 |
| **Mantle** | 79 | 79 | 79 |
| **Total** ($N \pm \sqrt{N}$) | 1025 ± 32 | 706 ± 27 | 175 ± 13 |
| **Expected Signal** | 79 ± 109 | 79 ± 84 | 79 ± 20 |

A measurement of the mantle signal $M$ requires subtracting the non-mantle background $B$ from the total $N$. The uncertainty is $\delta M = \delta N + \delta B$. This is shown graphically in Figure 1. Hanohano is capable of measuring the mantle U/Th neutrino flux to 25% in one year. After four years of operation, the 20% systematic uncertainty in background from the crust would begin to dominate, ultimately limiting the measurement at the 8% level. This same uncertainty severely limits the capability of detectors at continental sites for measuring the mantle flux. For example neither of the other detectors considered would make a positive detection from the same exposure. We note that it would take SNO+ and Borexino 15 and 50 years, respectively, to achieve this exposure. This result demonstrates the advantage of an oceanic site over continental sites for measuring the U/Th neutrino flux from Earth's mantle.

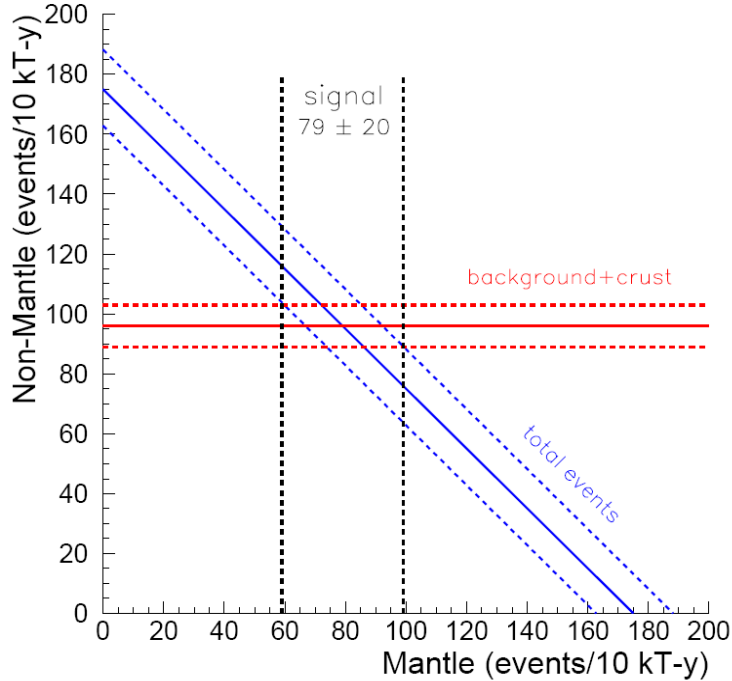

Figure 1. The background-subtracted mantle signal measured by Hanohano with a 10 kT-y exposure.

## 3. Geo-reactor Neutrino Detection Sensitivity

The energy spectrum of anti-neutrinos produced in a nuclear fission reactor extends from 0 MeV up to about 10 MeV, much wider than the 0 to 3.4 MeV for geo-neutrinos. In the energy region above the threshold energy for anti-neutrino interaction with target 1.8 MeV to 3.4 MeV, geo-neutrinos are a background to the geo-reactor. Therefore a lower energy threshold of 3.4 MeV is applied to the geo-reactor search to completely remove this background. We set an upper bound of 9.3 MeV for convenience (the probability that a geo-reactor neutrino has greater energy is negligible). In this energy range (3.4 – 9.3 MeV) the geo-reactor neutrino event rate is 38 events / $TW_t$ / 10 kT-y.

### 3.1 Geo-reactor Neutrino Background

The background to geo-reactor neutrino detection comes from the same sources as for geo-neutrino detection, but is lower at higher energies with the exception of the contribution due to lithium, which is perhaps a factor of two higher (Tolich 2005). The event rates differ because of the different energy window and different analysis cuts. We summarize them in Table 2. At a depth of 4 km the total background rate is 30 per 10 kT-y, compared to the signal for a 1 TW$_t$ geo-reactor of 38 per 10 kT-y. The signal detection significance for a 1 TW$_t$ geo-reactor from a 10 kT-y exposure is $S/\sqrt{S+B}$ = 4.6 sigma. If the geo-reactor exists at the high power end of predicted range, 10 TW$_t$, the detection significance would increase to 19 sigma. Assuming we have pinned down the background (via fitting the spectrum and other means), the geo-reactor power can be resolved to 5-22% going from the upper to lower expected power levels (10-1 TW$_t$), limited by statistical fluctuations, not systematic uncertainty.

Table 2: Background sources and corresponding rates expected for mid-ocean geo-reactor neutrino detection. We assume the $^{210}$Po background to be 100 times purer than the level for the scintillating oil at KamLAND and the reactor background to be known to 4%.

| Geo-reactor Background | Rate (10 kT-y)$^{-1}$ |
|---|---|
| $^9$Li (4 km) | 4 ± 1 |
| $^{210}$Po | 1 ± 1 |
| Accidentals | 1 ± 0 |
| Commercial Reactors | 24 ± 1 |
| **Total Background** | **30 ± 2** |
| **Geo-reactor Signal** | **38/TW$_t$** |

As with geo-neutrino detection, we performed simulation studies of Hanohano's sensitivity to the geo-reactor assuming various signal and background levels and an initial 10 kT-y exposure. Using the knowledge of expected spectrum for both signal and background will add to our confidence in detecting any geo-reactor signal, as well as improving background estimate precision. We confirm that the error on a positive power measurement will remain dominated by statistical fluctuations, not systematic uncertainty.

We have not herein considered the further confirmation of the location of any positive geo-neutrino signal by neutrino direction measurement. The direction to a neutrino source was measured by the Chooz experiment team (Apollonio et al., 2000) (operating nearby a power reactor) by using the difference between the location of the initial (positron annihilation) and second (neutron capture) effective vertices, and is due to the slight neutron momentum in the direction of the incoming neutrino. The prospects for confirmation that the putative geo-reactor signal is coming from generally the direction of the center of Earth depends upon detector design (electronics and photomultiplier time resolution and scintillating oil decay lifetime). Our first estimates are not very encouraging unless the geo-reactor is at the higher end of the possible power levels. Another possibility for increasing directionality involves loading the scintillating liquid to reduce the neutron capture time and make the radiation length shorter. This possibility requires study.

We conclude that the geo-reactor measurement is easier than the mantle geo-neutrino measurement. If the geo-reactor exists at the level suggested, (and important for driving plumes) we should be able to convincingly detect it and make useful measurements of the power.

## 4. Recommended Detector Specifications

Based on the foregoing results, we recommend the following for a deep ocean anti-neutrino observatory:

- A detector fiducial volume of about 10 kT (some 20 times KamLAND).

- The planned live time should be at least 1 year, yielding a 10 kT-y exposure at each location.

- A depth of 4 km is sufficient to comfortably accomplish the twin geophysics goals. Depths greater than about 4 km do not make significant improvements in background levels.

- The scintillating oil must be as free of $^{210}$Po as possible, with a goal of 100 times less than the initial KamLAND contamination. With testing at KamLAND already demonstrating reduction by a factor of $10^5$ using a newly developed distillation process this goal should be easily attainable.

- A deep ocean (or mid-ocean island) location, far from continents, is required to resolve the mantle flux of U/Th decay neutrinos from background including neutrinos from oceanic and continental crusts.

- A location near the Hawaiian Island land mass does not add significant crust background to the measurement of U/Th neutrinos from the mantle.

- With the stated goal of 10 kT-y exposure and 4 km depth and expected backgrounds we can achieve a 25% measurement of the U/Th neutrinos from the mantle, and hence this level of global concentration.

- Again, with stated assumptions on exposure and backgrounds, we can measure a geo-reactor power to 5-22% precision for source powers in the predicted range of 10-1 TW$_t$. With a null result, we can set upper limits to the power of <0.5 TW$_t$ at >95% confidence level.

## 5. Conclusions

In summary, we find that a 1 year deployment of a 10 kT, deep ocean anti-neutrino observatory can achieve the geophysics goals of this proposed experiment: a measurement of mantle geo-neutrinos and a definitive search for the hypothetical geo-reactor. We show that the 20% systematic uncertainty of the background from U/Th neutrinos in Earth's crusts prevents detectors at continental sites from measuring mantle geo-neutrinos and ultimately limits the measurement at an oceanic site to 8%. Subsequent deployments of Hanohano at other oceanic locations present the opportunity to measure lateral heterogeneity of U/Th concentrations in the mantle at the 25% level.

**Acknowledgements:** We are grateful to Robert Svoboda and Nikolai Tolich for many useful suggestions and comments. This work was partially funded by U.S. Department of Energy grant DE-FG02-04ER41291 and by the National Defense Center of Excellence for Research in Ocean Sciences (CEROS). CEROS is a part of the Natural Energy Laboratory of Hawaii Authority (NELHA), an agency of the Department of Business, Economic Development and Tourism, State of Hawaii. CEROS is funded by the Defense Advanced Research Projects Agency (DARPA) through grants and agreements with NELHA. This work does not necessarily reflect the position or policy of the Government, and no official endorsement should be inferred.

## References

Apollonio, M. *et al*.: 2000, Phys. Rev. D **61**, 012001.

Araki, T. *et al*.: 2005a, *Experimental investigation of geologically produced antineutrinos with KamLAND*, Nature (London) **436**, 499.

Araki, T. *et al*.: 2005b, Phys. Rev. Lett. **94**, 081801.

Avilez, C., Marx, G., and Fuentes, B.: 1981, *Earth as a source of antineutrinos*, Phys. Rev. D **23**, 1116.

Chen, M.C. (these proceedings).

Detwiler, *et al.*: 2002, *Nulcear Propelled Vessels and Neutrino Oscillation Experiments*, Phys. Rev. Lett. **89**, 191802.

Domogatski, G. *et al.*: 2004, *Neutrino Geophysics at Baksan I: Possible Detection of Georeactor Antineutrinos*, hep-ph/0401221.

Eder, G.: 1966, *Terrestrial Neutrinos*, Nucl. Phys. **78**, 657.

Enomoto, S., Ohtani, E., Inoue, K., and Suzuki, A.: 2005, hep-ph 0508049.

Enomoto, S.: 2005, Neutrino Geophysics and Observation of Geo-Neutrinos at KamLAND, Ph.D. Thesis, Tohoku University.

Fiorentini, G., Lissia, M., Mantovani, F., and Vannucci, R.: 2004, hep-ph/0401085.

Giammarchi, M.G. and Miramonti, L. (these proceedings).

Herndon, J.M.: 1996, Proc. Nat. Acad. Sci., **93**, 646.

Hollenbach, D.F. and Herndon, J.M.: 2001, Proc. Nat. Acad. Sci., **98**, 11085.

Krauss, L.M., Glashow, S.L., and Schramm, D.N.: 1984, *Antineutrino astronomy and geophysics*, Nature (London) **310**, 191.

Mantovani, F. *et al.*: 2004, Phys. Rev. D **69**, 013001.

Maricic, J.: 2005, *Setting Limits on the Power of a Geo-Reactor with KamLAND*, Ph.D. Thesis, University of Hawaii.

McDonough, W.F. and Sun, S. −s.: 1995, Chem. Geol. **120**, 223.

Mei, D.: 2006. *Muon-Induced Background Study for Underground Laboratories,* Phys.Rev. D **73**, 053004.

Pakvasa, S.: 2005, Nucl. Phys. B (proceedings supplement) **145**, 378.

Raghavan, R.S. *et al*.: 1998, *Measuring the Global Radioactivity in the Earth by Multidetector Antineutrino Spectroscopy*, Phys. Rev. Lett. **80**, 635.

Raghavan, R.S.: 2002, *Detecting a Nuclear Fission Reactor at the Center of the Earth,* hep-ex/0208038.

Rothschild, C.G., Chen, M.C., and Calaprice, F.P.: 1998, *Antineutrino geophysics with liquid scintillator detectors*, Geophys. Res. Lett. **25**, 103.

Suzuki, A. (these proceedings).

Tolich, N.: 2005, Experimental Study of Terrestrial Electron Anti-neutrinos with KamLAND, Ph.D. Thesis, Stanford University.